\documentclass[12pt]{article}
\usepackage{latexsym}
\usepackage{axodraw}
\usepackage{epsfig}
\usepackage{amsfonts}
\usepackage{array,longtable}
\begin{document}

 \title{\bf
Renormalization Group Treatment of Nonrenormalizable Interactions}
\author{D.I.Kazakov$^{1,2,\dag}$ and G.S.Vartanov$^{1,3,\ddag}$}
\date{}
\maketitle \vspace{-0.8cm}
\begin{center}
{\it $^1$Bogoliubov Laboratory of Theoretical Physics, Joint
Institute for Nuclear Research, Dubna, Russia \\[0.1cm]
$^2$Institute for Theoretical and Experimental Physics, Moscow,
Russia\\[0.1cm]$^{3}$Moscow Institute of Physics and Technology, Moscow, Russia}
\end{center}

\begin{abstract}
The structure of the UV divergencies in  higher dimensional
nonrenormalizable theories is analysed. Based on renormalization
operation and renormalization group theory it is shown that even
in this case  the leading divergencies (asymptotics) are governed
by the one-loop diagrams the number of which, however, is
infinite. Explicit expression for the one-loop counter term in an
arbitrary D-dimensional quantum field theory without derivatives
is suggested. This allows one  to sum up the leading asymptotics
which are independent of the arbitrariness in subtraction of
higher order operators. Diagrammatic calculations in a number of
scalar models in higher loops are performed to be in agreement
with the above statements. These results do not support the idea
of the na\"ive power-law running of couplings in nonrenormalizable
theories and fail (with one exception) to reveal any simple closed
formula for the leading terms.
\end{abstract}

\section{Introduction}

It has been commonly accepted that one cannot use
nonrenormalizable interactions beyond the tree level due to
uncontrollable ultraviolet divergencies. Nothing has changed in
understanding of this problem; however, these days it is sometimes
suggested that nonrenormalizable interactions are treated on equal
footing with the renormalizable ones. This specially became
fashionable within the context of extra dimensional theories, all
of them being nonrenormalizable in a usual sense. A wide spread
opinion ensures these theories to be treated as effective
ones~\cite{Rattazzi}, meaning that one believes that the UV
troubles are cured somehow by including of these theories in a
more general framework, for instance, a string theory while
considering the low energy effective action. In the latter case,
one distinguishes the "relevant", "marginal", and "irrelevant"
operators a'la Wilson~\cite{Wilson}, so that at low energies one
can abandon contributions from irrelevant operators being power
suppressed and end up with relevant and marginal operators which
are renormalizable. \vfill
\hrule\vskip 1mm\noindent {$^{\dag}$ e-mail: kazakovd@theor.jinr.ru \\
$^{\ddag}$ e-mail: vartanov@theor.jinr.ru} \\
However, this is not the case of higher dimensional theories since
there are no relevant or  marginal operators there. They are all
irrelevant or nonrenormalizable and one cannot throw away all of
them. Hence, one has to find the way to deal with them or to give
up.

 Evidently there are no problems at the tree level, where one can
drop the contributions from the higher order operators, but in the
loops one is faced with severe properties of nonrenormalizable
interactions: appearance of an infinite sequence of higher order
operators. This is true even at low energies (smaller than an
intrinsic scale set up by dimensional couplings).

Sometimes one talks about "low energy renormalizability" of
nonrenormalizable theories assuming the  ignorance of  the
contributions of these higher order operators. Within this context
one discusses the "power-law running" of couplings~\cite{Dienes}.
This seemingly an attractive idea needs a thorough investigation.

In higher dimensional theories (meaning the dimension is higher
than the critical dimension of a given interaction which is
usually 4) one can approach this problem in two ways: either to
consider the theory directly in (flat) extra dimensions and try to
cope with power-like divergencies, or to use the kind of
Kaluza-Klein approach with (compact) extra dimensions and consider
4-dimensional theory with an infinite tower of K-K
modes~\cite{Giudice}. In the first case, one faces the problem of
appearance of new higher dimensional operators as the UV counter
terms; and in the latter one, the problem of divergence of the
sums over the K-K states. Nevertheless, it is claimed that at low
energies when we ignore (in a sense of effective theory) the
higher order operators or cuts the divergence of the sum
introducing a cut-off momentum, we get the power-like behaviour of
the Green functions or the power-law running of the original
couplings~\cite{Dienes}. Similar results follow from the
renormalization group approach a'la Wilson based on the
$\epsilon$-expansion and  analytical continuation of perturbation
theory expansion above the critical dimension. In this case, one
has a nonperturbative fixed point where the theory possesses
conformal invariance so that the effective coupling becomes
dimensionless that is sometimes referred to nonperturbative
renormalizability~\cite{K1,Morris}. Here, the higher order
operators are suppressed in the infrared domain and the Green
functions in the vicinity of the fixed point exhibit the
power-like behaviour~\cite{K2}.

However, the running literally means the summation of the leading
asymptotics (the leading logs or leading powers).  Hence, assuming
these considerations to be correct, one sums up the leading
contributions of an infinite sequence of diagrams into the
"running" quantities. If this is true, at least at low energies,
one has to be able to check by explicit calculations of diagrams
how the leading terms are summed up. This means that they are
essentially predicted prior to the calculation. It is very well
known how it works in renormalizable theories within the
renormalization group approach. The question is whether it also
works in the nonrenormalizable case.

The purpose of this paper is to demonstrate that indeed the
structure of local quantum field theory even in nonrenormalizable
theories reveals the one-loop origin of the leading asymptotics
despite the fact that there is no simple closed expression for the
amplitudes like in the renormalizable case. Our results, including
explicit calculation of the diagrams in some nonrenormalizable
models, do not support the idea of the na\"ive "power-law running"
of couplings and lead (with one miraculous exception) to
complicated expressions without obvious summation pattern.

\section{Renormalization operation in local QFT}

The essence of the diagram behaviour can be figured out from the
structure of the so-called R(enormalization)-operation valid in
any local QFT. It basically states that genuine UV divergencies
even in nonrenormalizable theories are always local (or
quasi-local), i.e. contain a limited number of derivatives. So the
bare Lagrangian being properly regularized contains all possible
local counter terms. In their turn, these counter terms are in
one-to-one correspondence with the logarithms that appear on top
of the powers of momenta.

Since the bare Lagrangian does not depend on the renormalization
scale, the explicit dependence of the counter terms on the scale
has to be compensated by the inexplicit dependence of the
couplings. In the case of a renormalizable Lagrangian, there is
one or a few couplings, and differentiating the bare Lagrangian
with respect to a scale one gets the corresponding RG equations.
For the dimensional regularization one gets the so-called pole
equations~\cite{poles} that relate the higher order poles with the
lowest one in all orders of perturbation theory. In particular,
taking the one-loop contribution to the lowest pole one recovers
the whole infinite series of the leading pole terms. In the case
of a single coupling they are summed into a geometric progression
\begin{equation}\label{coupling}
  g^{bare}=(\mu^2)^\varepsilon \frac{g}{1+bg/\varepsilon},
\end{equation}
where the coefficient $b$ comes from the one-loop $\beta$-function.  For several
couplings the expressions are more complicated but the higher order terms can still  be
summed up and are completely defined by the one-loop contribution.

For the nonrenormalizable Lagrangian, the R-operation still holds,
but now one has an infinite number of terms in the bare
Lagrangian. Even if one starts with the finite number of terms new
higher order operators will be generated via the UV counter terms.
Moreover, since the couplings in the nonrenormalizable case are
dimensionfull (as it always happens in extra dimensions), the
diagrams reveal only higher dimensional operators to compensate
the negative dimension of the coupling. The original operators do
not appear. So the counter terms possess  the triangle structure:
each operator gives a contribution to the renormalization of the
higher operators and not to itself and lower ones. Still, despite
this cumbersome picture, it has the same structure as in
renormalizable theories. The higher order poles are still defined
by the lowest one and the leading ones are determined by the
one-loop diagrams. The corresponding pole equations for a general
QFT theory were written in Ref.\cite{general}. We present them
here without a derivation
\begin{eqnarray}
{\cal L}^{Bare} &=&  (\mu^2)^\varepsilon ({\cal L} + \sum_{n=0}^\infty
\frac{A_n({\cal L})}{\varepsilon^n}),\\
({\cal L}\frac{\delta}{\delta {\cal L}}-1)A_n({\cal L})&=&\beta({\cal L})
\frac{\delta}{\delta {\cal L}}A_{n-1}({\cal L}), \ \beta({\cal L})=({\cal
L}\frac{\delta}{\delta {\cal L}}-1)A_1({\cal L}),\label{poles}
\end{eqnarray}
where $A_{n}({\cal L})$ means that the corresponding counter term is calculated starting
from the Lagrangian ${\cal L}$.

Performing loop expansion and taking into account that the counter terms $A_{n}({\cal
L})$ are homogeneous functions
$$A_n({\cal L})=\sum_{k=n}^\infty A_{nk}({\cal L}), \ \ \ A_{nk}(\lambda{\cal
L})=\lambda^k A_{nk}({\cal L}), $$ (here the first subscript
denotes the order of the pole term; and the second one, the number
of loops) one can reduce eq.(\ref{poles}) to
\begin{eqnarray}\label{pole}
  (\lambda\frac{\delta}{\delta\lambda}-1)A_n(\lambda{\cal L})|_{\lambda
  =1}&=&\frac{d}{d\eta}A_{n-1}({\cal L}+\eta\beta({\cal L}))|_{\eta =0}\\
  \beta({\cal L})=(\lambda\frac{\delta}{\delta\lambda}-1)A_1(\lambda{\cal L})|_{\lambda
  =1}&=&\sum_{k=n}^\infty kA_{1k}({\cal L}).
\end{eqnarray}
 For the leading poles this leads to
\begin{eqnarray}
\ A_{nn}({\cal L})&=& \frac 1n \ \frac{d}{d\eta}A_{n-1\ n-1}({\cal L}+\eta A_{11}({\cal
L}))|_{\eta=0},\label{rec}
\end{eqnarray}

One can clearly see from eq.(\ref{rec}) that if one knows the
one-loop contribution to the simple pole, namely $A_{11}$, than
one knows via a recursive procedure all the leading terms
$A_{nn}$. We would like to stress ones again that this statement
and eq.(\ref{rec}) are true in any theory, renormalizable or
nonrenormalizable.

However, there is some crucial difference in application of this
recursion to renormalizable or nonrenormalizable theories. While
in the renormalizable theories one has a finite number of one-loop
diagrams contributing to $A_{11}$ that are constructed from the
original Lagrangian, in the nonrenormalizable case in general one
has an infinite number of such one-loop diagrams involving new
higher dimensional operators. So unless one has some hint how
these diagrams are related to one another, one has an infinite
number of unknown coefficients in $A_{11}$.

Consider the scalar field theories when the original interaction Lagrangian does not
contain derivatives. In four dimensions the explicit closed expression for $A_{11}$ has
the form~\cite{algorithm}
\begin{equation}\label{d4}
A_{11}({\cal L}) = -\frac{1}{(4\pi)^2}\frac{1}{4}\ \
  \frac{\delta^2{\cal L}}{\delta\phi^2}\times
  \frac{\delta^2{\cal L}}{\delta\phi^2}.
\end{equation}
Natural generalization of this formula to D-dimensions would be
\begin{equation}\label{dd}
A_{11}({\cal L}) =
-\frac{1}{(4\pi)^{D/2}}\frac{\Gamma(D/2-1)}{4\Gamma(D-2)}\ \
  \frac{\delta^2{\cal L}}{\delta\phi^2}
  (\partial^2)^{D/2-2}
  \frac{\delta^2{\cal L}}{\delta\phi^2}.
\end{equation}
However, when $D>4$ there are other one-loop divergent diagrams
having more external legs. With taking into account higher
dimensional operators which appear in higher loops the number of
these terms increases.  To reproduce this increasing sequence of
terms we conjecture that in $D$ dimensions the one-loop counter
term has the form
\begin{eqnarray}\label{11}
  A_{11}({\cal L}) &=& -\frac{1}{(4\pi)^{D/2}}\frac{\Gamma(D/2-1)}{4\Gamma(D-2)}\ \
  \frac{\delta^2{\cal L}}{\delta\phi^2}
  \frac{(\partial^2)^{D/2}}{(\partial^2+\frac{\delta^2{\cal L}}{\delta\phi^2})^2}
  \frac{\delta^2{\cal L}}{\delta\phi^2}\nonumber\\
  &=& -\frac{1}{(4\pi)^{D/2}}\frac{\Gamma(D/2-1)}{4\Gamma(D-2)}\ \ \frac{\delta^2{\cal L}}{\delta\phi^2}
  \frac{(\partial^2)^{D/2-2}}{(1+\partial^{-2}\frac{\delta^2{\cal L}}{\delta\phi^2})^2}
  \frac{\delta^2{\cal L}}{\delta\phi^2},
\end{eqnarray}
where the denominator is understood as a geometric progression
with derivatives acting on expansion terms so as to cancel all
nonlocalities $\partial^{-2}$. Hence, for a given $D$ one has only
a finite number of terms to contribute. However, while calculating
$A_{nn}$, according to eq.(\ref{rec}), one has to replace ${\cal
L}$ by $A_{11}$, which may contain extra derivatives. These extra
derivatives can also cancel the $\partial^{-2}$ terms , so the
expansion goes further. The {\it general recipe} is: use
eq.(\ref{rec}) for $A_{nn}$ with $A_{11}({\cal L)}$ given by
eq.(\ref{11}) and expand the denominator until the nonlocal terms
are cancelled by the corresponding derivatives. Surely, at a given
order of perturbation theory only a finite number of expansion
terms contribute.

 Below we illustrate these statements considering calculation of the leading
order terms explicitly in a number of models. These calculations prove that
eq.(\ref{rec}) is valid, but the numbers obtained (with one exception) do not reveal any
obvious summation pattern.

\section{Explicit calculations in nonrenormalizable models}

For the sake of simplicity we consider a set of massless scalar
field theories within the framework of dimensional regularization.
When the coupling $g$ in a given dimension $D$ has a negative
dimension,  the theory is nonrenormalizable in the usual sense.

\subsection{D=6, $\mathbf{\phi}^4_{(6)}$}

Let us start with the Lagrangian
\begin{equation}\label{l6}
  {\cal L}_{int} = -\frac{\lambda\phi^4}{4!}
\end{equation}
in $D=6-2\varepsilon$. The coupling $\lambda$ has a negative
dimension $[\lambda]=-2+2\varepsilon$ that leads to an infinite
series of the counter terms containing higher order operators
generated via loop expansion. One has, according to eq.(\ref{11}),
\begin{equation}
A_{11}({\cal L}) = -\frac{1}{(4\pi)^{3}}\frac{1}{24}\ {\cal L}"
  \frac{\partial^2}{(1+\partial^{-2}{\cal L}")^2}{\cal L}"
=-\frac{1}{(4\pi)^{3}}\frac{1}{24}\left(\frac{\lambda^2}{4} \phi^2 \partial^2 \phi^2+
\frac{\lambda^3}{4} (\phi^2)^3\right)\label{one},
\end{equation}
where following the above-mentioned recipe we omitted the nonlocal
terms. Hereafter we use the notation ${\cal L}" \equiv
\frac{\delta^2{\cal L}}{\delta\phi^2}$. Two terms of expansion in
eq.(\ref{one}) correspond to the one-loop two-point and triangle
diagrams, respectively.

If one substitutes eq.(\ref{11}) into (\ref{rec}), one gets for
$A_{22}$
\begin{eqnarray}\label{22}
 2A_{22}({\cal L}) &=& -\frac{1}{(4\pi)^{3}}\frac{1}{24}\left\{ [A_{11}({\cal L})]"
  \frac{\partial^2}{(1+\partial^{-2}{\cal L}")^2}{\cal L}"
-2{\cal L}"\frac{[A_{11}({\cal L}]"}{(1+\partial^{-2}{\cal
L}")^3}{\cal L}" \right. \nonumber\\
&+& \left.{\cal L}"\frac{\partial^2}{(1+\partial^{-2}{\cal
L}")^2}[A_{11}({\cal L})]"\right\},
\end{eqnarray}
where $A_{11}$ is given by eq.(\ref{one}). Substituting
(\ref{one}) into (\ref{22}) and performing expansion of the
geometric progression, one finally gets
\begin{eqnarray}\label{two}
A_{22}&=&-\left(\frac{1}{(4\pi)^{3}}\frac{1}{24}\right)^2\left\{\frac{\lambda^3}{8}
(\phi^2\partial^2 \phi^2)"\partial^2 \phi^2 +\frac{\lambda^4}{8}[(\phi^2)^3]"\partial^2
\phi^2+\frac{3\lambda^4}{16}(\phi^2\partial^2 \phi^2)"(\phi^2)^2 \right. \nonumber \\
&+& \left. \frac{3\lambda^5}{16}[(\phi^2)^3]"(\phi^2)^2 +
\frac{3\lambda^5}{16}(\phi^2)^2(\phi^2\partial^2 \phi^2)"\partial^{-2} \phi^2\right\}.
\end{eqnarray}

The tricky point here is the interference of the variational derivative with respect to
$\phi$ and the ordinary space-time derivative. This requires some definition. Let us
first evaluate the space-time derivatives and then the variational derivatives
\begin{eqnarray}\label{der}
(\phi^2\partial^2 \phi^2)"& = &   2(\phi^3\partial^2\phi +
\phi^2(\partial\phi)^2)"  \nonumber \\
&=&   4\left( 3\phi\partial^2\phi + 3\phi^2\partial^2 + 4\phi\partial\phi\partial +
(\partial\phi)^2 + \phi^2\partial\partial \right).
\end{eqnarray}
When the space-time derivative is left free, we  understand it as
acting on the propagator of the corresponding one-loop diagram.
Schematically it is shown below 
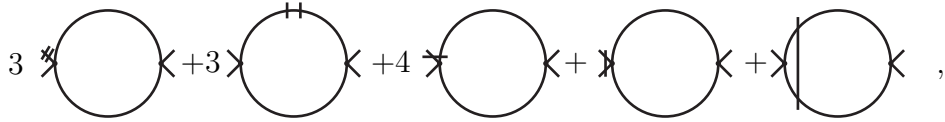
\vspace{0.2cm}

\begin{center}
\begin{picture}(260,20)(0,0)\SetWidth{2}\SetScale{0.5}
\Text(-35,0)[]{3}\Oval(0,0)(40,40)(0) \Vertex(40,0){2}\Vertex(-40,0){2}
\Line(-50,10)(-40,0) \Line(-50,-10)(-40,0) \Line(50,10)(40,0) \Line(50,-10)(40,0)
\SetWidth{2} \Line(-47,1)(-41,11) \Line(-50,3)(-44,13)  \Text(35,0)[]{+3}
 \Oval(140,0)(40,40)(0)
\Vertex(100,0){2} \Vertex(180,0){2} \Line(90,10)(100,0) \Line(90,-10)(100,0)
\Line(190,10)(180,0) \Line(190,-10)(180,0) \SetWidth{2} \Line(145,34)(145,46)
\Line(135,34)(135,46) \Text(107,0)[]{+4}
 \Oval(290,0)(40,40)(0) \Vertex(330,0){2}
\Vertex(250,0){2} \Line(240,10)(250,0) \Line(240,-10)(250,0)
\Line(340,10)(330,0) \Line(340,-10)(330,0) \SetWidth{2}
\Line(237,5)(256,5) \Text(177,0)[]{+} \Oval(420,0)(40,40)(0)
\Vertex(460,0){2} \Vertex(380,0){2} \Line(370,10)(380,0)
\Line(370,-10)(380,0) \Line(470,10)(460,0) \Line(470,-10)(460,0)
\SetWidth{2} \Line(375,10)(375,-10) \Text(245,0)[]{+}
\Oval(550,0)(40,40)(0) \Vertex(590,0){2} \Vertex(510,0){2}
\Line(500,10)(510,0) \Line(500,-10)(510,0) \Line(600,10)(590,0)
\Line(600,-10)(590,0) \SetWidth{2} \Line(520,35)(520,-35)
\Text(315,-4)[]{,}
\end{picture}
\end{center}\vspace{0.4cm}
where the crossed lines mean the corresponding derivatives.

The next step is the reduction of the diagram with derivative(s)
to the one without them. Usually this step is straightforward and
is performed by analysing the corresponding one-loop integral. In
this particular case, one has
\begin{eqnarray}
(\phi^2\partial^2 \phi^2)"\begin{picture}(30,20)(10,-5) \SetScale{0.3}\SetWidth{2}
\Oval(100,0)(40,40)(0)\Line(50,10)(60,0) \Line(50,-10)(60,0) \Line(150,10)(140,0)
\Line(150,-10)(140,0)\end{picture}& \Rightarrow &   4\left(
3p_1^2+3*0-\frac{4pp_1}{2}+p_1p_2 +\frac{p^2}{2}\right)
\phi\phi  \nonumber \\
&=& 8p_1^2\phi\phi=8 \phi\partial^2\phi,
\end{eqnarray}
where $p$ is the momentum entering the loop, $p_1$ and $p_2$ are
the momenta of each leg. Here we take the symmetrical point where
$p^2=\frac{4}{3}p_1^2=\frac{4}{3}p_2^2$, $p_1p_2 =
-\frac{1}{3}p_1^2$.

Now let us take the third term in (\ref{two}). Here, we have
triangle diagrams

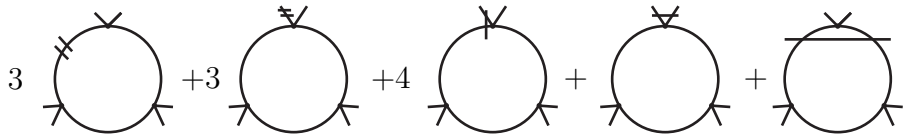
\begin{center}
\begin{picture}(260,20)(0,0) \SetScale{0.5}\SetWidth{2}
\Text(-35,0)[]{3} \Oval(0,0)(40,40)(0)  \Vertex(0,40){2} \Vertex(35,-20){2}
\Vertex(-35,-20){2} \Line(10,50)(0,40) \Line(-10,50)(0,40) \Line(10,50)(0,40)
\Line(49,-21)(35,-20) \Line(42,-35)(35,-20) \Line(-49,-21)(-35,-20)
\Line(-42,-35)(-35,-20)  \Line(-40,25)(-30,15) \Line(-37,32)(-27,21) \Text(35,0)[]{+3}
\Oval(140,0)(40,40)(0)  \Vertex(140,40){2} \Vertex(175,-20){2} \Vertex(105,-20){2}
 \Line(130,55)(140,40) \Line(150,55)(140,40) \Line(189,-21)(175,-20)
\Line(182,-35)(175,-20) \Line(91,-21)(105,-20) \Line(98,-35)(105,-20)
\Line(128,52)(138,52) \Line(130,48)(140,48) \Text(107,0)[]{+4} \Oval(290,0)(40,40)(0)
\Vertex(290,40){2} \Vertex(325,-20){2} \Vertex(255,-20){2} \Line(300,55)(290,40)
\Line(280,55)(290,40) \Line(339,-21)(325,-20) \Line(332,-35)(325,-20)
\Line(241,-21)(255,-20) \Line(248,-35)(255,-20) \Line(285,30)(285,52) \Text(177,0)[]{+}
\Oval(420,0)(40,40)(0) \Vertex(420,40){2} \Vertex(455,-20){2} \Vertex(385,-20){2}
\Line(410,55)(420,40) \Line(430,55)(420,40) \Line(469,-21)(455,-20)
\Line(462,-35)(455,-20) \Line(371,-21)(385,-20) \Line(378,-35)(385,-20)
\Line(410,48)(430,48) \Text(245,0)[]{+} \Oval(550,0)(40,40)(0) \Vertex(550,40){2}
\Vertex(585,-20){2} \Vertex(515,-20){2} \Line(560,50)(550,40) \Line(540,50)(550,40)
\Line(560,50)(550,40) \Line(599,-21)(585,-20) \Line(592,-35)(585,-20)
\Line(501,-21)(515,-20) \Line(508,-35)(515,-20) \Line(510,30)(590,30)
\end{picture}
\end{center}\vspace{0.5cm}

Again reducing the diagram with derivatives to the one without them one obtains
\begin{eqnarray}
(\phi^2\partial^2 \phi^2)" \begin{picture}(30,20)(10,-5) \SetScale{0.3}\SetWidth{2}
\Oval(100,0)(40,40)(0)\Line(110,50)(100,40) \Line(90,50)(100,40) \Line(149,-21)(135,-20)
\Line(142,-35)(135,-20) \Line(51,-21)(65,-20) \Line(58,-35)(65,-20)
\end{picture}& \Rightarrow & 4\left( 3(-\frac{p_{12}^2}{3}) + 3p_1^2
-\frac{8}{5}p_1^2 +\frac{5}{6}p_{12}^2 \frac{pp_1}{2} \right)\phi\phi=
\frac{7\lambda^4}{10}\phi\partial^2\phi, \nonumber
\end{eqnarray}
where $p_1$ and $p_2$ are the momenta in each leg and $p_{12}$ is the momentum entering
in one vertex. Here we take the symmetrical point where $p_1p_2=-\frac{1}{5}p_1^2$ and
$p_{12}^2 = \frac{8}{5} p_1^2 $.

We are left with the last term in eq.(\ref{two}). Here one has the box diagram which is
convergent in D=6 if there are no derivatives. This means that only the second and the
last terms of eq.(\ref{der}) contribute. One has

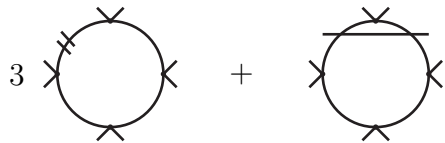
\begin{center}
\begin{picture}(260,40)(0,-20) \SetScale{0.5}\SetWidth{2}
\Text(65,0)[]{3} \Oval(200,0)(40,40)(2) \Vertex(200,40){2}  \Vertex(200,-40){2}
\Vertex(160,0){2} \Vertex(240,0){2} \Line(210,50)(200,40) \Line(190,50)(200,40)
\Line(210,-50)(200,-40) \Line(190,-50)(200,-40) \Line(250,10)(240,0)
\Line(250,-10)(240,0) \Line(150,10)(160,0) \Line(150,-10)(160,0)  \Line(160,25)(170,15)
\Line(163,32)(173,21)
 \Text(150,0)[]{+} \Oval(400,0)(40,40)(2)
\Vertex(400,40){2} \Vertex(400,-40){2} \Vertex(360,0){2}
\Vertex(440,0){2} \Line(410,50)(400,40) \Line(390,50)(400,40)
\Line(410,-50)(400,-40) \Line(390,-50)(400,-40)
\Line(450,10)(440,0) \Line(450,-10)(440,0) \Line(350,10)(360,0)
\Line(350,-10)(360,0) \Line(360,30)(440,30)
\end{picture}
\end{center}
The first diagram is a triangle one and the second is easily reduced to it. As a result
one has
\begin{eqnarray}
(\phi^2\partial^2 \phi^2)" \begin{picture}(30,20)(10,-5) \SetScale{0.3}\SetWidth{2}
\Oval(100,0)(40,40)(0) \Line(110,50)(100,40) \Line(90,50)(100,40) \Line(110,-50)(100,-40)
\Line(90,-50)(100,-40) \Line(150,10)(140,0) \Line(150,-10)(140,0) \Line(50,10)(60,0)
\Line(50,-10)(60,0)
\end{picture}& \Rightarrow &   4\left( \frac 32 - \frac 12\right)(-)
\phi\phi = -4\phi\phi,
\end{eqnarray}
where the minus sign of the last term is  due to the reduction of
the box diagram to a triangle one.

Adding up all terms eq.(\ref{two}) finally leads to
\begin{eqnarray}\label{red}
A_{22}&=&-\left(\frac{1}{(4\pi)^{3}}\frac{1}{24}\right)^2\left\{{\lambda^3}
(\phi\partial^2 \phi)\partial^2 \phi^2 +\frac{15\lambda^4}{4}(\phi^2)^2\partial^2
\phi^2+\frac{7\lambda^4}{10}\phi\partial^2 \phi (\phi^2)^2 \right. \nonumber \\
&& \left.\hspace*{3cm} + \frac{45\lambda^5}{8}(\phi^2)^2(\phi^2)^2 -
\frac{3\lambda^5}{4}(\phi^2)^2(\phi^2)^2\right\}.
\end{eqnarray}

One may proceed further and get
\begin{eqnarray}\label{33}
  A_{33}&=&-\left(\frac{1}{(4\pi)^{3}}\frac{1}{24}\right)^3\left\{\frac{\lambda^4}{6}\left[
2(\phi\partial^2 \phi)\partial^2(\phi\partial^2 \phi)+((\phi\partial^2
\phi)\partial^2\phi^2)"\partial^2\phi^2 \right. \right. \nonumber
\\ &&\left. \left. \hspace*{3.7cm} +(\phi^2\partial^2(\phi\partial^2\phi))"\partial^2\phi^2\right]+
... \right\} = \\
&=&-\left(\frac{1}{(4\pi)^{3}}\frac{1}{24}\right)^3\left\{\frac{2\lambda^4}{3}\left[
2(\phi\partial^2 \phi)\partial^2(\phi\partial^2
\phi)+\partial^2\phi\partial^2\phi\partial^2\phi^2+\phi\partial^2\partial^2\phi\partial^2\phi^2\right]+
... \right\}\nonumber,
\end{eqnarray}
where we keep only the terms quartic in fields.

To compare these expressions with explicit diagram calculation, it
is useful to transfer to the momentum representation. Then the
space-time derivative means some momenta with the proper
symmetrization which depends on the number of legs. Thus,
eqs.(\ref{one},\ref{red}) and (\ref{33}) become
\begin{eqnarray}\label{mom}
A_{11}&=& -\frac{1}{(4\pi)^{3}}\frac{1}{24}\left(-\frac{\lambda^2}{12}
(\phi^2)^2(s+t+u) + \frac{\lambda^3}{4} (\phi^2)^3\right), \nonumber \\
A_{22}&=&-\left(\frac{1}{(4\pi)^{3}}\frac{1}{24}\right)^2\left\{\frac{\lambda^3}{12}
(s+t+u)^2 (\phi^2)^2 -\lambda^4\frac{ 67}{60}(\phi^2)^3\sum_{i=1}^6 p_i^2
+ \lambda^5\frac{39}{8}(\phi^2)^4\right\}, \nonumber \\
A_{33}&=&-\left(\frac{1}{(4\pi)^{3}}\frac{1}{24}\right)^3\left\{-\frac{\lambda^4}{72}\left[
2(\phi^2)^2 +(\phi^2)^2+(\phi^2)^2\right](s+t+u)^3+ ... \right\}.
\end{eqnarray}
We have checked by explicit diagram calculations that
eqs.(\ref{mom}) are indeed correct.

It is interesting to consider the 4-point function. The
Lagrangian, together with the corresponding counter terms, look
like
 \begin{equation}\label{4}{\large
\begin{array}{l} \phi^4 :
 -\frac{\lambda}{24}+\frac{\lambda^2}{(4\pi)^3}\frac{1}{24\varepsilon}\frac{s+t+u}{12}-
  \frac{\lambda^3}{(4\pi)^6}\frac{1}{(24\varepsilon)^2}\frac{(s+t+u)^2}{12}+
  \frac{\lambda^4}{(4\pi)^9}\frac{1}{(24\varepsilon)^3}\frac{(s+t+u)^3}{72}+ ...  \\
  =-\frac{\lambda}{24}\left[1-\frac{\lambda}{(4\pi)^3\varepsilon}\frac{s+t+u}{12}+
(\frac{\lambda}{(4\pi)^3\varepsilon})^2(\frac{s+t+u}{12})^2\frac 12 -
(\frac{\lambda}{(4\pi)^3\varepsilon})^3(\frac{s+t+u}{12})^3\frac 16 + ...\right] \\
= -\frac{\lambda}{24}\exp(-\frac{\lambda}{(4\pi)^3\varepsilon}\frac{s+t+u}{12}) \ \ \ \
?!
\end{array}}
\end{equation}

One can see that the first three terms remarkably remind the
expansion of the exponent. We have checked this fact by {\it
explicit} calculation of the diagrams up to {\it four} loops and
with the help of the {\it pole equations} (\ref{poles}) up to
$A_{55}$. In the latter case, unlike the diagram calculation it is
straightforward and can be easily continued further. So far we
failed to find all order proof that everything is summed up to the
exponent, though it seems quite reasonable.

The pole relation (\ref{4}) evidently leads to the corresponding expression for the
four-point function
\begin{equation}\label{4p}
  \Gamma_4=\exp[\frac{\lambda}{(4\pi)^3}\frac{s}{4}\log(s/\mu^2)],
\end{equation}
where we substituted symmetric asymptotics $s=t=u$.

Two comments are in order:\\
1) In nonrenormalizable theories, due to the presence of an
infinite series of operators one has an infinite number of
normalization conditions. This is why the theory is not defined.
However, the leading poles (or the leading logarithms) are
independent (!) of
these conditions. Thus, the leading behaviour is defined unambiguously.\\
2) Equation (\ref{4p}) has quite an unusual form different from
the geometric progression expected from the na\"ive "power-law
running". Is it occasional  or a common feature of
nonrenormalizable theories needs to be clarified.

To check this exponential behavior we made similar calculations in several other models.

\subsection{$\phi^4_{(8)},\ \phi^4_{(10)}, \ \& \ \phi^4_{(D)}$}

Let us consider first the $D=8$ case. According to eq.(\ref{11}),
$A_{11}$ takes the form
\begin{eqnarray}
A_{11}({\cal L}) &=& -\frac{1}{(4\pi)^{4}}\frac{1}{240}\ {\cal L}"
  \frac{\partial^4}{(1+\partial^{-2}{\cal L}")^2}{\cal L}" \nonumber \\
&=& -\frac{1}{(4\pi)^{4}}\frac{1}{240}\left(\frac{\lambda^2}{4} \phi^2 \partial^4 \phi^2+
\frac{\lambda^3}{4} \partial^2\phi^2(\phi^2)^2 +
\frac{3\lambda^4}{16}(\phi^2)^4\right)\label{8},
\end{eqnarray}
where we again omitted the nonlocal terms. Obviously, with
increase of dimension the length of the $A_{11}$ term increases;
$A_{22}$ then becomes
\begin{eqnarray}
  A_{22}&=&-\left(\frac{1}{(4\pi)^{4}}\frac{1}{240}\right)^2\left\{
  \frac{\lambda^2}{8}(\phi^2\partial^4\phi^2)"\partial^4\phi^2 +...\right\},
\end{eqnarray}
where we kept only the terms quartic in fields.

Here again we are faced with the problem of evaluating the
variational derivatives. We first evaluate the space time
derivatives
\begin{eqnarray}
(\phi^2\partial^4\phi^2)"&=&4\partial^2\phi\partial^2\phi+4\phi\partial^2\partial^2\phi+16\partial\phi\partial
\partial^2\phi+8\partial\partial\phi\partial\partial\phi \nonumber \\
&+&8\phi(2\partial^2\phi\partial^2+\partial^2\partial^2\phi+\phi\partial^2\partial^2+4\partial\partial^2\phi\partial
+4\partial\phi\partial\partial^2+4\partial\partial\phi\partial\partial) \nonumber \\
&+& 2\phi^2(4\partial\partial\ \partial\partial).
\end{eqnarray}
The meaning of partial derivatives acting on the right is understood as above in a sense
of acting on the lines of the one-loop diagrams. This gives
\begin{eqnarray}
 A_{22}&=&
  -\left(\frac{1}{(4\pi)^{4}}\frac{1}{240}\right)^2 \times \label{228} \\ &\times& \left\{\frac{\lambda^2}{8}
\left(\frac{20}{7}\partial^2\phi\partial^2\phi+4\phi\partial^2\partial^2\phi+2\partial^2\partial^2\phi^2
+\frac{120}{7}\partial\partial\phi\partial\partial\phi+16\partial\phi\partial\partial^2\phi\right)
  \partial^4\phi^2+ ...\right\},\nonumber
\end{eqnarray}

Unlike the D=6 case, expression for $A_{22}$ in D=8 does not look
simple. Indeed, in the momentum representation it is not expressed
in terms of the Mandelstam variables $s,t$ and $u$  and at the
symmetrical point looks like
 \begin{eqnarray}\label{44}
& \phi^4 :&
 -\frac{\lambda}{24}-\frac{\lambda^2}{(4\pi)^4}\frac{1}{240\varepsilon}\frac{s^2}{4}-
  \frac{\lambda^3}{(4\pi)^8}\frac{1}{(240\varepsilon)^2}\frac{55s^4}{112}+... \nonumber\\
  &=& -\frac{\lambda}{24}\left(1+\frac{\lambda}{(4\pi)^4}\frac{s^2}{40\varepsilon}+
  \frac{\lambda^2}{(4\pi)^8}\frac{s^4}{(40\varepsilon)^2}\frac{55}{168}+ ... \right).
\end{eqnarray}
This result is also confirmed by explicit diagrammatic
calculation. One cannot see any trace of exponent here. Moreover,
it does not look like any other simple function since, as we have
already mentioned, it cannot be expressed in terms of the
Mandelstam variables.

We have carried out the same calculation in D=10.
Equation(\ref{44}) in this case becomes
 \begin{eqnarray}\label{410}
& \phi^4 :&
 -\frac{\lambda}{24}+\frac{\lambda^2}{(4\pi)^5}\frac{1}{3360\varepsilon}\frac{s^3}{4}-
  \frac{\lambda^3}{(4\pi)^{10}}\frac{1}{(3360\varepsilon)^2}\frac{91s^6}{192}+... \nonumber\\
  &=& -\frac{\lambda}{24}\left(1-\frac{\lambda}{(4\pi)^5}\frac{s^3}{560\varepsilon}+
  \frac{\lambda^2}{(4\pi)^{10}}\frac{s^6}{(560\varepsilon)^2}\frac{91}{288}+ ... \right).
\end{eqnarray}
Again one can see no trace of exponent. To understand it better,
we have calculated the second order term in the case of $\phi^4$
theory in $D$ dimensions. The simplest way it can be done is by
explicit diagram calculation. The result is
\begin{eqnarray}\label{4D}
 \phi^4 :
   &-&\frac{\lambda}{24}\left\{1+\frac{\lambda}{(4\pi)^{D/2}}\frac{(-1)^{D/2}3\Gamma(D/2-1)}{2\Gamma(D-2)}
   \frac{s^{D/2-2}}{\varepsilon}\right. \nonumber  \\
   &+&\left. \left(\frac{\lambda}{(4\pi)^{D/2}}\frac{(-1)^{D/2}3\Gamma(D/2-1)}{2\Gamma(D-2)}
   \frac{s^{D/2-2}}{\varepsilon}\right)^2 c(D)
  + ... \right\},
\end{eqnarray}
where
$$c(D)=\frac 13 +
\frac{2^{5-D}}{3^{3-D/2}}F_{21}(\frac{3D}{2}-4,2-\frac{D}{2};\frac{D-1}{2}|\frac
13).$$ This complicated expression is remarkably simplified for
particular values of $D$ and at low dimensions is:
\begin{center}
\begin{tabular}{|l|l|l|l|l|l|l|}\hline c & D=4 & D=6 & D=8 & D=10 & D=12& D $\to\infty$\\ \hline  & 1
&1/2 &55/168 & 91/288& 6005/18304 & $\to$ 1/3\\
\hline
\end{tabular}
\end{center}


\section{Conclusion}

Summarizing we would like to stress once again that
\begin{enumerate}
\item Direct calculations of Feynman diagrams demonstrate all characteristic features and
problems of nonrenormalizable interactions;
\item For nonrenormalizable interactions like for renormalizable ones the leading
divergences (asymptotics) are defined by the one-loop diagrams;
\item In the nonrenormalizable case contrary to the renormalizable one the number of
one-loop diagrams is infinite;
\item We conjectured the form of the one-loop counter term in arbitrary scalar QFT with
derivativeless interactions and checked it by explicit calculations in lower loops;
\item Using the renormalization group technique, it is possible  to sum up the leading asymptotics
which are independent of the arbitrariness in subtraction of higher order operators;
\item Unlike the renormalizable case, this summation does not reveal (with one exception so far)
any simple function;
\item The na\"ive power-law running seems not to be valid.
\end{enumerate}

\section*{Acknowledgements}
Financial support from RFBR grant \# 05-02-17603 and grant of the Ministry of Science and
Technology Policy of the Russian Federation \# 2339.2003.2 is kindly acknowledged. G.V.
would like to thank Dynasty Foundation for support. We are grateful to S.Mikhailov for
valuable discussions.

\end{document}